# Ultra Violet radiation from sunlight: A key moderating factor in the spread of SARS-CoV-2 during the pandemic of 2020

Nirmala D. Desouza[1], Srikanta Sannigrahi[*2], D. Blaise[3], Daniel Tan[4], Artemi Cerdà[5]

[1]Department of Physics, Visvesvarayya National Institute of Technology, Nagpur, Maharashtra, India; nirmala_devidas@yahoo.co.in

[2]School of Architecture, Planning and Environmental Policy, University College Dublin, Ireland srikanta.sannigrahi@ucd.ie

[3]Department of Crop Production, ICAR-Central Institute for Cotton Research, Nagpur, Maharashtra, India; blaise_123@rediffmail.com

[4]The University of Sydney, School of Life and Environmental Sciences, Faculty of Science, New South Wales, Sydney, Australia; daniel.tan@sydney.edu.au

[5]University of Valencia, Spain (artemio@uv.es)

*Corresponding author: Srikanta Sannigrahi (srikanta.sannigrahi@ucd.ie)

**Abstract**

Lockdowns imposed in most of the countries were lifted following a decline in the COVID-19 cases towards May-June 2020. A recent surge (second wave) in the COVID-19 cases in Europe and other temperate countries as compared to the tropical regions suggests the likely role of solar radiation. We hypothesized that ultraviolet radiation's effect might be a significant factor moderating the spread of the COVID-19 across countries. Regression analysis was done for the UV radiation data for seven hotspot cities (New Delhi, Mumbai, Milan, Madrid, New York, Melbourne and Sydney) with the daily COVID-19 cases. Global erythemal UV radiation values were lower during winter and higher during summer. In general, the daily new COVID-19 cases registered were higher during the winter months having low UV radiation dose (0.5-3.7 kJ $m^{-2}$). Cases began to decline with summer onset that corresponded to increased UV radiation (2.5-6.7 kJ $m^{-2}$). Our studies suggest that the natural UV radiation could be a strong determining factor moderating the spread of COVID-19 cases. The importance of UV radiation in natural sunlight as a disinfectant for SARS-CoV-2 cannot be ignored since the lockdowns were lifted; further, it can be considered as a factor for COVID management.

## 1. Introduction

The COVID-19 pandemic caused by the novel coronavirus, SARS-COV-2, was first detected in Wuhan, China, in December 2019. As of 23 February 2021, there have been 111,419,939 confirmed cases of COVID-19, including 2,470,772 deaths (WHO, 2021). Among the countries, the United States of America, Brazil, Mexico, India have been affected most severely by COVID-19 (Maiti et al., 2021; Chakraborti et al., 2020). Getting infected with the coronavirus is mainly due to either coming in close contact with an infected person and/or contact with contaminated fomites or surfaces (van Doremalen et al., 2020) or through aerosols (Mittal et al., 2020; Morawska and Cao, 2020). All these studies have emphasized the disease spread in-doors. Most nations imposed nationwide lockdowns to restrict travel and coming in contact with infected people (Sannigrahi et al., 2020a, 2020b, Basu et al., 2020; Kutela et al., 2021; Mollalo et al., 2021). Further, the wearing of masks was made mandatory in most countries. After the number of cases began to decline, lockdowns were lifted, and people were seen moving freely without masks or wearing them without adequately covering their mouth and/or nose. In such situations, there is every likelihood that the virus may get transmitted through the droplets/aerosols (Mittal et al., 2020) in the out-door environment. The assumption that the risk of COVID-19 transmission in out-door environments is low (Chan et al., 2020) appears to be a risky approach, especially because of the breach in social distancing norms and gathering in large numbers after the lockdown was lifted. Furthermore, small respiratory droplets have been reported to travel distances as far as tens of metres (Feng et al., 2020; Morawska and Cao, 2020).

Weihs et al. (2012) identified the influence of solar radiation, especially ultraviolet (UV) radiation, on the infectivity of the closely related SARS virus. This is so because the wavelength range suitable for the inactivation of influenza viruses is 100-380 nm, which is in the UV radiation range. The entire UV radiation range is divided into 3 bands: *UVA (315-400* *nm), UVB (280-315 nm) and UVC (100-280 nm).* Sunlight reaching the Earth's surface does not contain significant irradiance for wavelengths less than 290 nm because of absorption by atmospheric ozone, increased Rayleigh scattering and decreasing wavelength. Thus, the entire UVC and 90% of UVB wavelength are absorbed by the atmosphere, and hence, UV radiation reaching the surface of the Earth is mostly UVA and small amounts of UVB (WHO, 2020). Fortunately, the primary photochemical processes that damage the viral DNA or RNA occurs at all the solar UV radiation wavelengths, varying only inefficiency at different wavelengths (Rauth, 1965). The interaction between the UV radiation and DNA or RNA can lead to the destruction of cells' replication ability (Tseng and Li., 2005). Simulated sunlight inactivated 90% of SARS-CoV-2 every 6.8 minutes in simulated saliva and every 14.4 minutes in culture media (Ratnesar-Shumate et al., 2020). Carvalho et al. (2020) reported that solar UV radiation has a high potential to inactivate coronaviruses, but the degree strongly depends on the location as well as the season. There is abundant solar radiation during the summer as compared to the winter season. Further, sunlight irradiance is greater in the tropics as compared to the sub-tropics and temperate regions. Thus, we cannot ignore this aspect of the natural sunlight that may play a major part in containing the disease spread. Few studies have considered this aspect probably due to the WHO's basic assumption and emphasis that transmission is mainly through contact indoors by contact with infected people (Chan et al., 2020).

We hypothesized that the rapid increase in the COVID-19 cases with the onset of winter and greater severity in the sub-tropics and temperate regions was possibly due to low UV radiation received while the spread is slower in the tropics that had an abundance of solar radiation. Our main objective was to understand the influence of the UV in the real world and determine if there was a relationship between the UV radiation received and the daily COVID-19 cases before the lockdown, during lockdown and post lockdown.

## 2. Methodology

### *2.1. Locations*

Seven cities were chosen for the study to represent cities that were the major hotspots of the novel coronavirus across latitudes (**Table 1**) to verify the effect of solar radiation on the novel coronavirus. Major cities in the sub-tropics include New Delhi, Milan, Madrid, New York and Sydney, while Melbourne has a temperate climate and Mumbai was selected to represent the tropics.

### *2.2. Data used*

Data on the number of COVID-19 cases on a daily basis were collected for the different cities from their respective Public Health Agencies based on PCR diagnosis (**Table 1**). To assess the extent to which UV radiation received from solar radiation could effectively contain the spread of COVID-19, we obtained data on the UV radiation from the Tropospheric Emission Monitoring Internet Service (TEMIS) archives (Van Geffen et al., 2018). TEMIS, formerly part of the Data User Programme of the European Space Agency, is a web-based service to browse and download atmospheric satellite data products such as tropospheric trace gases, aerosol concentration as well as UV radiation data and surface albedo. Daily data from TEMIS were taken for the cloud-free erythemal UV radiation for New Delhi, Mumbai, New York, Sydney and Melbourne. Cloud-modified erythemal UV radiation data were used for Milan and Madrid.

To understand any relationship of UV radiation, we performed regression analyses of the daily COVID-19 cases with UV radiation values and calculated Pearson's correlation coefficient. The Scatter plot of the daily UV radiation with the daily new COVID-19 cases showed a non-linear relationship. Therefore, we performed regression by adopting the distributed lag non-linear model approach (Guo et al., 2019). Regression analysis was performed in three time periods: before lockdown, during the lockdown, and post-lockdown.

Regressions (polynomial or exponential) that gave the best fit were considered because of their simplicity and predictive ability (Thomas and Finney, 1984). Statistical analysis was done using MYSTAT and a p-value <0.05 was considered as significant.

## 3. Results

### *3.1. Daily COVID-19 cases across geographies*

After the first reports of the COVID-19 case in Wuhan, China, in December 2019, it was reported in more than 180 countries by the end of March 2020. Confirmed cases of COVID-19 ranged from 25,515 (Milan) to as high as 102,863 cases (New York) in the first month after the first COVID-19 case was reported. The spread of COVID-19 cases was rapid across Europe and North America, with peak occurrences recorded during March-April 2020 (**Fig. 1**). Although COVID-19 cases were also reported in the tropics, the increase in the number of cases was slow, with 2,095 confirmed cases in Mumbai and 268 in New Delhi. The rise in the COVID-19 cases began to decline in mid-May in all the cities of the sub-tropics and temperate regions in this study. A second wave is being observed with the rising number of cases since October (and continues to rise). Following the first reports of the imported COVID-19 cases into Australia on 1 March 2020, COVID-19 cases in Sydney and Melbourne showed a steep rise in April followed by a decline in the month of May. However, a second wave in the COVID-19 cases was reported during July-August 2020 in Melbourne (Fig. 1).

### *3.2. Daily UV radiation*

A closer look at the UV radiation data indicates 2.4 to 4.7-fold higher values of UV radiation received in the tropics compared with the sub-tropical and temperate regions. In general, the entire tropical zone had high values of the erythemal UV radiation with a decline in values as we moved to higher latitudes (**Fig. 1**).The sub-tropical and temperate regions (Europe and North America) received low UV radiation during September-April, with the UV radiation ranging from 0.5-1.9 kJ $m^{-2}$. The sub-tropical cities of Milan, Madrid and New York in the northern hemisphere had UV dose <2 kJ $m^{-2}$ on all the days of January, February, October and November. At a higher latitude, Milan (45° N) had begun to record values <2 kJ $m^{-2}$ by mid-September. Mumbai (18.9° N) had values >2 kJ $m^{-2}$right through the year. The UV radiation increased with the approaching summer season (April-May), resulting in a bell-shaped curve. Mean values of the UV radiation received during May to August ranged from 3.1-7.3 kJ $m^{-2}$. In the Indian cities, the mean values of the UV radiation received during the summer season ranged from 4.9-6.1 kJ $m^{-2}$. For the countries lying in the southern hemisphere, the UV radiation values declined with time (February-March to July-August). This is evident from the data for Sydney and Melbourne in Australia (**Fig. 1**). UV radiation continued to increase from August, giving rise to a U-shaped curve.

### *3.3. Daily New COVID-19 cases vs UV radiation*

Scatter plots of the daily new COVID-19 cases vs the erythemal UV dose for the seven locations included in this study showed a significant relationship (**Fig. 1a-g, Fig. 2, Fig. 3**). In general, the daily COVID-19 cases were inversely related to UV radiation. The best fit relationship with the UV radiation was explained either by a second-order polynomial or an exponential equation. An exponential decline in the COVID-19 cases was observed with an increase in the UV radiation in Milan and Melbourne. A second-order polynomial best described the relationship of the COVID-19 cases and UV for New Delhi, Mumbai, Madrid, New York and Sydney (**Table 2**). Among the cities, the association between UV dose and COVID counts significantly varied with time (**Fig. 3**). In Madrid and New York, the positive association (during the pre-lockdown phase) between these two factors turned to be negative during the lockdown and post-lockdown period (**Fig. 3**). While in some cities, i.e. Mumbai and New Delhi, the association was mainly positive during the pre-lockdown and lockdown period and became negative once the lockdown restriction is lifted (**Fig. 3**). Inverse associations among the parameters (UV dose and COVID counts) were observed during the first pre-lockdown phase in Australian cities, i.e. Sydney and Melbourne, which is not common for the other cities in this study.

## 4. Discussion

*The association between UV radiation and COVID incidents*

Daily UV radiation is the total amount of UV radiation integrated between sunrise and sunset, accounting for the variation in solar zenith angle and cloud cover fraction (Holick et al., 2005). High UV radiation over the tropical regions is because of the greater monthly average irradiance exposure since the sun is directly overhead (Herman, 2010). This explains the latitudinal effect of the UV irradiance at various places with high UV radiation in the tropics and moderate values in the sub-tropics, and low values in the temperate regions during the period of the study. Further, variations were also seen within countries with the cities lying to the north (New Delhi vs Mumbai) having lower values to those in the south. In the southern hemisphere, Melbourne had lower UV radiation in winter than Sydney as Melbourne is closer to the South Pole. Significant correlation observed between the COVID-19 cases with UV radiation suggests that the high UV radiation received in the tropics could be a probable reason for the slow spread of the COVID-19 initially (**Fig. 1**) despite having the world's most densely populated cities. Furthermore, most tropical countries in Africa, Asia, and Latin America (not analyzed in this study for the lack of TEMIS data) have reported fewer COVID-19 cases. For instance, Bangkok was the first to report a COVID-19 case outside of Wuhan, China. However, the total number of cases in the entire Thailand was just 25,599 (as of 23 February 2021), probably due to high levels of UV radiation that may have affected the virulence and infectivity of the virus (Sangripanti and Lytle, 2007). The low UV radiation in the sub-tropics and temperate regions resulted in a rapid spread of COVID-19, and the daily cases began to increase rapidly. A drop in the COVID-19 cases began to occur following a rise in the UV radiation. With the commencement of summer, the UV radiation received began to increase between 23.5

$^{o}$N and 23.5 $^{o}$S and was possibly one reason for a decline in the number of COVID-19 cases over Wuhan, China and the sub-tropical and temperate countries. Carvalho et al. (2020) showed the strong potential of solar radiation in inactivating the virus depends on the location. These results support our findings with fewer COVID-19 cases reported at the lower latitudes while greater cases are being reported at higher latitudes, such as those in Milan, Madrid, New York and Melbourne. Although UV radiation damages the viral RNA (Rauth, 1965), values lower than 2 kJ $m^{-2}$ may not be as effective in containing the spread of the disease. At higher latitudes (>40$^{o}$) of the sub-tropics and the temperate regions, UV radiation <2 kJ $m^{-2}$ was recorded on > 120 days in a year while they were ~38 days at 23.5$^{o}$ S and negligible at 18$^{o}$ N. Interestingly, in Melbourne, the number of daily COVID-19 cases began to peak in July-August with the UV radiation on a decline in the southern hemisphere as the countries started to move into the winter season. After the first peak of COVID-19 cases in April, the disease was brought effectively under control for two main reasons: (i) effective social distancing norms being followed and (ii) a clean unpolluted air environment. Nevertheless, the substantial spike reported in July was reportedly due to breaches in the hotel quarantine in Melbourne, exacerbated by poor contact tracing, followed by a very hard lockdown. In addition, the lower amounts of UV radiation in winter might have contributed to the spread.

*Confounding effects of factors on UV radiation availability and resulted impact on COVID counts*

Our results suggest the role of the UV radiation in containing the infectious nature of the virion because of the strong and significant relationship of the COVID-19 cases with the UV radiation. UV exposure results in sufficient vitamin D levels in the body that can be stored in the body fat for subsequent utilization (Holick, 2007). This explains the lagged effects of UV radiation. Adequate vitamin D may have contributed to the reduced severity of COVID-19 cases (Laird et al., 2020) in the tropics compared with temperate regions. During the lockdown period, COVID-19 cases continued to increase in some cities such as Mumbai, New Delhi, Sydney and Melbourne when the UV radiation was high. This could be due to a combination of factors such as (i) people confined indoors did not receive the benefit of the UV radiation, (ii) socio-demographic factors (Sannigrahi et al., 2020) and (iii) high salt sprays in coastal cities (Sydney, Melbourne, Mumbai) and the occurrence of dust storms (Heβling et al., 2020) in New Delhi. Although the UV irradiance was high in the Indian cities during the summer months, extremely cloudy weather probably attenuated UV radiation and vitamin D dose (produced in skin in sunlight) received on the earth's surface in Mumbai and New Delhi as well as other cities (not reported here). Salt sprays and dust apart from pollutants arising from vehicular traffic and heavy industries in these cities also may have contributed to high aerosol optical thickness. This could be a factor that might have led to an attenuation of the UV radiation (Deng et al., 2012). Ozone also absorbs UV radiation, preventing it from reaching the ground surface (Koronakis et al., 2002). Furthermore, violation of quarantine and social distancing is another factor that led to sudden spike in cases in the Australian cities (e.g., Melbourne which experienced a second lockdown). The efficiency of UV radiation deactivation of the virus depends not only upon the solar irradiation but also upon the host immunity resulting in the seasonality of the disease. Increased vitamin D synthesis in the skin due to increased UV radiation increases the likelihood of immunity and thereby reduces further infections. Vitamin D follows three different mechanisms, namely, (i) physical barrier, (ii) cellular natural immunity and (iii) adaptive immunity; in reducing the risk of viral infection and mortality. Further, vitamin D also hinders the replication of the virus (Bouillon et al., 2019). However, if a person is already infected and developed severe complications, UV radiation may not make a real difference (Moozhipurath et al., 2020). Therefore, the role of vitamin D is primarily protective and cannot serve as a substitute for clinical treatment (Hsiang et al., 2020). However, we suggest that the UV radiation and vitamin D made in the skin in sunlight cannot be discounted and ignored as people gather for out-door activities (beaches, parks, open markets etc.).

As the UV radiation declines during the winter season, people should be strongly advised to wear masks while they are outdoors, especially in the northern hemisphere. Special attention is needed in cities with heavy industrial activity and deteriorating air quality as reported in cities such as Mumbai and New Delhi; since UV radiation gets attenuated. Apart from the lower values in the UV dose, other meteorological factors and prolonged exposure to high $PM_{2.5}$ may have contributed to a rise in the COVID-19 cases among the immune-compromised people (Ikram et al., 2015). Studies are needed immediately to capture the virion in the atmosphere and understand the role of the atmospheric factors on the virion.

The time-varying association (both positive and negative) between UV radiation and COVID counts were studied across the cities. The analysis yielded noticeable evidence that the time factor is considerably regulating the association and could have varied effects of UV dose on COVID spread in cities. Among the other drivers, climate factors are considered the most important confounding factors and may substantially influence the disproportionate variation of virus spread. Additionally, a distinct pattern of associations is observed in cities belonging to different climate zones. For example, cities from temperate climates such as Milan, Madrid, and New York exhibited different association patterns between UV radiation and COVID cases than cities from tropical countries. The availability of shortwave solar radiation and seasonal variation of sunlight could be another reason for the opposite pattern of association among the cities. The starting and ending date of lockdown in different cities is another factor that might be responsible for the discrepancy observed among the response and control variables in studied cities.

## 5. Conclusions

Our study demonstrated a strong relationship between solar UV radiation and COVID-19 spread across latitudes and longitudes. In general, COVID-19 cases declined with increasing UV radiation. Furthermore, people confined indoors due to the lockdowns initially may be another factor for not obtaining desired benefits of UV radiation in the form of vitamin D. Therefore, in a clean unpolluted air environment, potential beneficial effects of UV radiation from solar radiation can be realized. Maintaining social distance and wearing masks are essential as the UV radiation declines with the approaching winter season resulting in fresh waves of COVID-19 outbreaks such as those in the cities across India as well those in the northern hemisphere.

## Acknowledgements

We are thankful to the Royal Netherlands Meteorological Society, ESA, for the Erythemal UV radiation global maps. We also acknowledge the Ministry of Health and Family Welfare, Government of India (mygov.in) for the daily COVID-19 cases in the Indian cities and the Public Health Agencies, and WHO for the data of New York, USA; Madrid, Spain; Milan and Venice, Italy; London, UK; Melbourne and Sydney, Australia.

## Figure citations

**Fig. 1** Temporal variation of the UV radiation and the number of new daily COVID-19 cases across seven cities, i.e. (a) Madrid, (b) Milan, (c) New York, (d) Mumbai, (e) New Delhi, (f) Melbourne, and (g) Sydney.

**Fig. 2** Relationship of the daily COVID-19 cases and the UV radiation prior to the lockdown, during the lockdown, and after the lockdown in Madrid, Milan, New York, Mumbai, New Delhi, Melbourne and Sydney (NS=non significant, *$p<0.05$; **$p<0.01$, ***$p<0.001$)

**Fig. 3** Association between UV radiation and COVID cases during pre-lockdown, during the lockdown, and after lockdown in Madrid, Milan, New York, Mumbai, New Delhi, Melbourne and Sydney

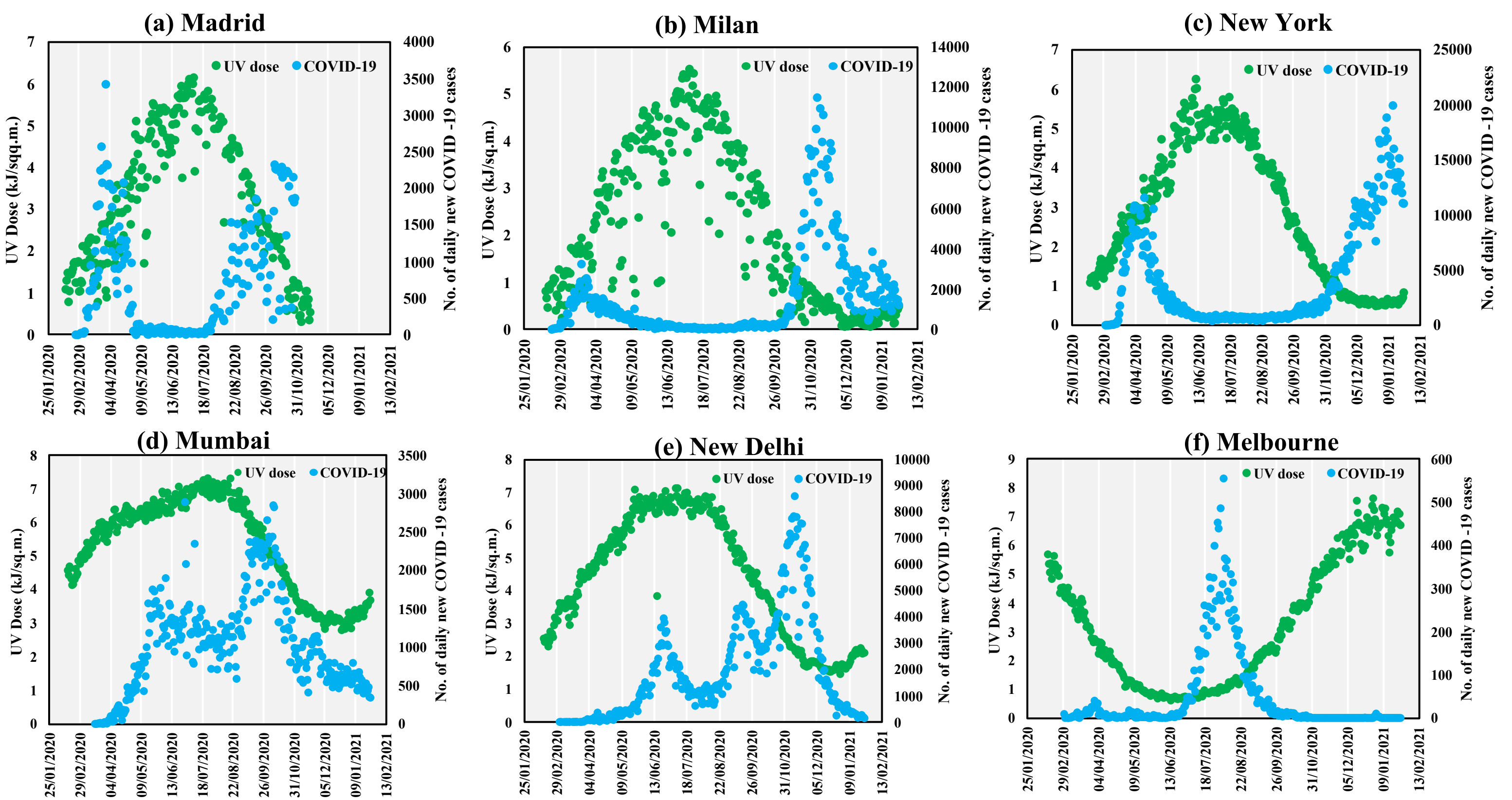

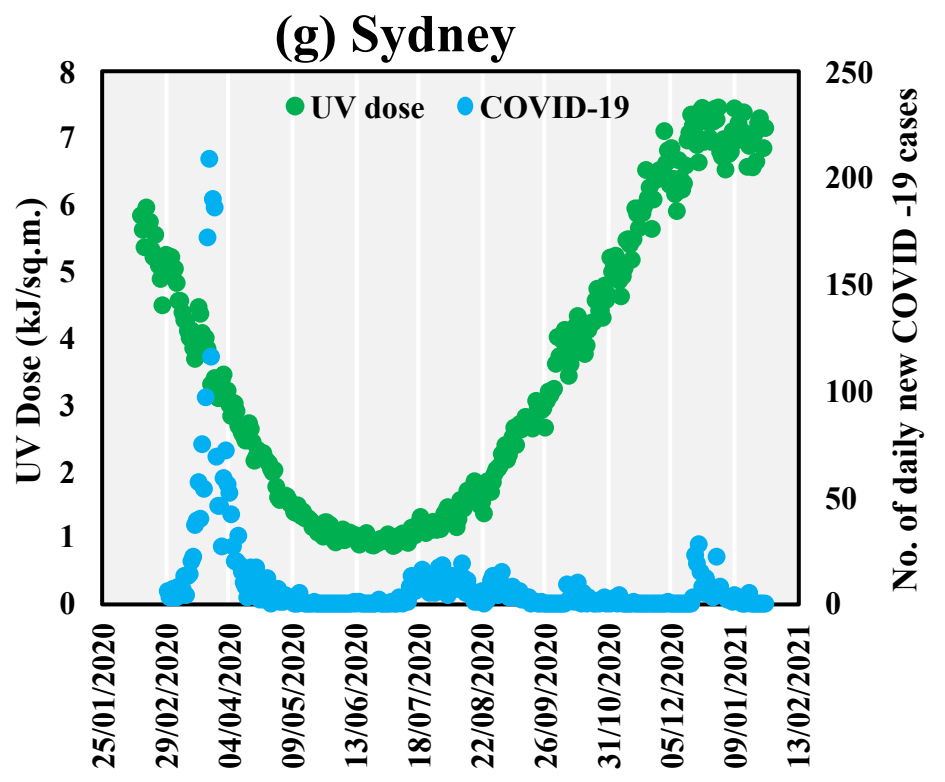

**Fig. 1**Temporal variation of the UV radiation and the number of new daily COVID-19 cases across seven cities, i.e. Madrid, Milan, New York, Mumbai, New Delhi, Melbourne, and Sydney.

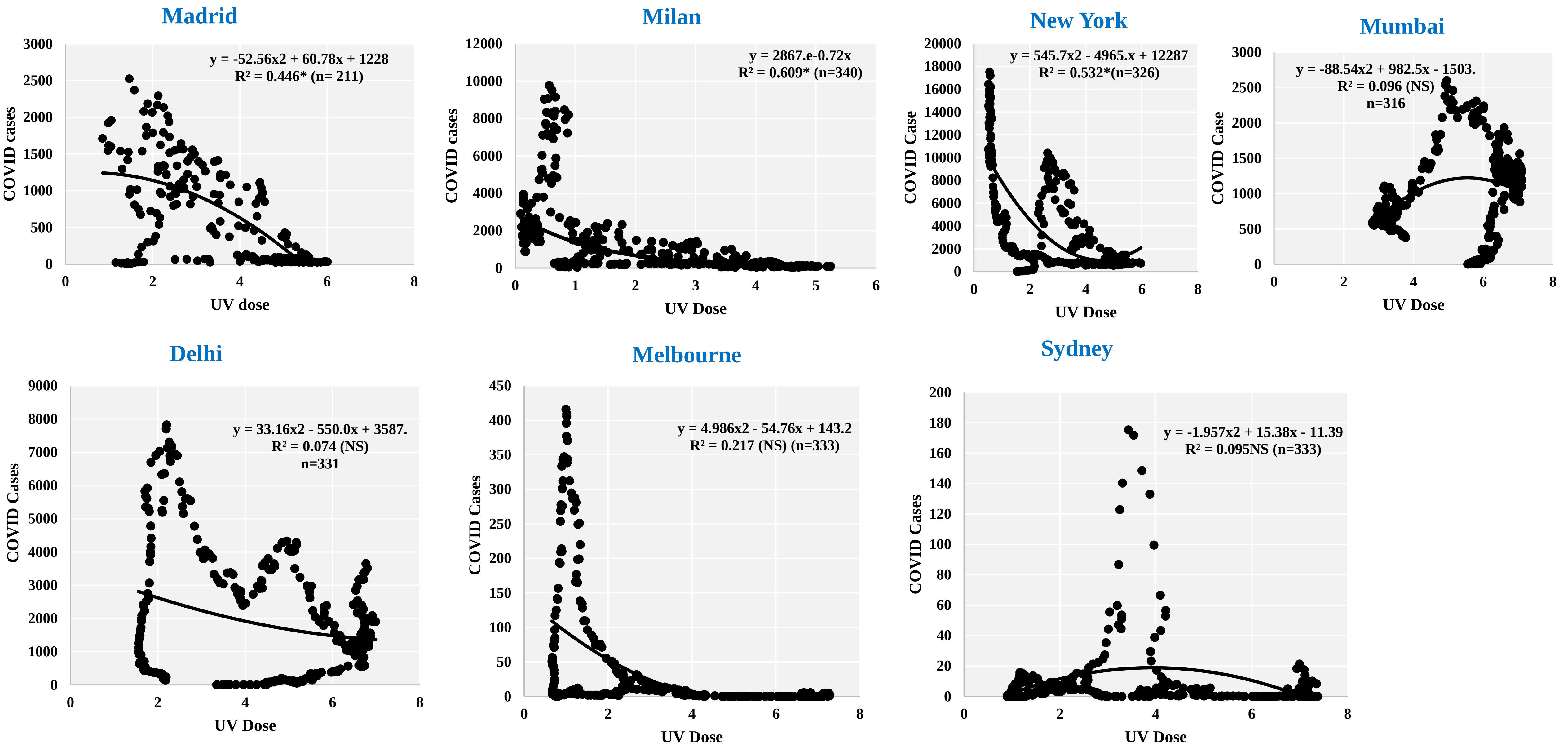

**Fig. 2** Relationship of the daily COVID-19 cases and the UV radiation prior to the lockdown, during the lockdown, and after the lock down in Madrid, Milan, New York, Mumbai, New Delhi, Melbourne and Sydney (NS=non significant, *$p<0.05$; **$p<0.01$, ***$p<0.001$)

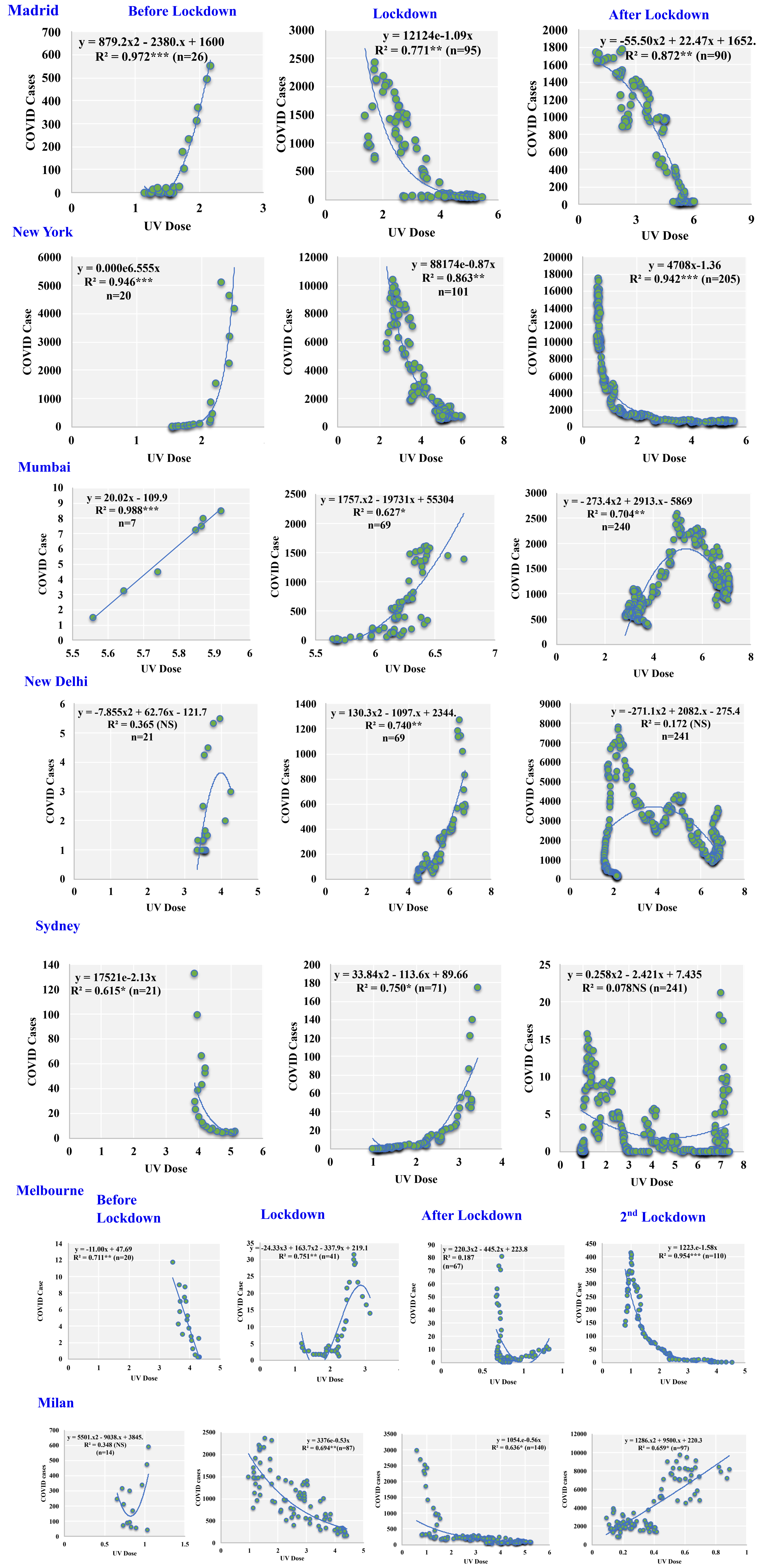

**Fig. 3** Association between UV radiation and COVID cases during pre-lockdown, during the lockdown, and after lockdown in Madrid, Milan, New York, Mumbai, New Delhi, Melbourne and Sydney.

**Table 1** Study locations for the different locations

| Location | Country | Lat Long | Reference |
|---|---|---|---|
| New Delhi | India | 28.22° N, 77.22° E | Ministry of Health and Family Welfare, Government of India (http://corona.mygov.in) |
| Mumbai | India | 18.93° N, 72.85° E | Ministry of Health and Family Welfare, Government of India (http://corona.mygov.in) |
| Milan | Italy | 45.81° N, 8.63° E | Protezione Civile (http://www.protezionecivile.gov.it/) |
| Madrid | Spain | 40.42° N, 3.7° W | Ministry of Health, Madrid |
| New York | USA | 40.64° N, 74.31° W | https://covid19tracker.health.ny.gov/views/NYS-COVID19-Tracker/NYSDOHCOVID-19Tracker-Map?%3Aembed=yes&%3Atoolbar=no&%3Atabs=n |
| Melbourne | Australia | 37.73° S, 145.1° E | https://www.covid19data.com.au/ |
| Sydney | Australia | 34.04° S,151.1 ° E | https://www.covid19data.com.au/ |

**Table 2** Regression equations and correlation coefficient ($R^2$) of the daily COVID-19 cases with the UV radiation.

| Location | UV dose (kJ $m^{-2}$) |
|---|---|
| New Delhi, India | $y = 33.16x^2 - 550.0x + 3587$ ($R^2 = 0.074^{NS}$; n=331) |
| Mumbai, India | $y = -88.54x^2 + 982.5x – 1503$ ($R^2 = 0.096^{NS}$; n=316) |
| Milan, Italy | $y = 2867.e^{-0.72x}$ ($R^2 = 0.609$*; n=340) |
| Madrid, Spain | $y = -52.56x^2 + 60.78x + 1228$ ($R^2 = 0.446$*; n= 211) |
| New York, USA | $y = 545.7x^2 - 4965.x + 12287$ ($R^2 = 0.532$*; n=326) |
| Melbourne, Australia | $y = 4.986x^2 - 54.76x + 143.2$ ($R^2 = 0.217^{NS}$; n=333) |
| Sydney, Australia | $y = -1.957x^2 + 15.38x - 11.39$ ($R^2 = 0.095^{NS;}$ n=333) |

*Significant at p<0.05; ** p<0.01, ***p<0.001